\documentclass[a4paper,10pt]{article}
\usepackage{amsmath}
\usepackage{amsfonts}
\usepackage{amssymb}
\usepackage{pstricks}
\usepackage{subfigure}
\usepackage{graphicx}
\usepackage{psfrag}
\usepackage{tikz}

\newcommand{\eref}[1]{Eq.~(\ref{#1})}

\newcommand{\nn}{\nonumber}

\def\refpos#1 #2 #3{\global\xrefpos=#1 \global\yrefpos=#2
                         \rlap{$\smash{#3}$}}
\def\put #1 #2 #3{\xput=#1 \yput=#2
                  \advance\xput by -\xrefpos
                  \advance\yput by -\yrefpos
                  \rlap{\kern\the\xput truebp
                        \vbox to 0pt{\vss\hbox{$\displaystyle #3$}
                        \kern\the\yput truebp}}}
\def\beginlabels\refpos#1\endlabels{\hbox{$\refpos#1$}}

\usepackage{amsfonts,amssymb}

\def\be{\begin{equation}}
\def\ee{\end{equation}}

\ifx\du\undefined
  \newlength{\du}
\fi
\begin{document}
\title{Energy scales in a holographic black hole and conductivity at finite momentum}
\author{Pallab Basu \\
        \small{ pallab@phas.ubc.ca} \\
 	\small{University of British Columbia} \\
	\small{Vancouver} \\
	\small{Canada, V6T 1Z1}}
\maketitle

\begin{abstract}

In this work we discuss the low temperature ($T$) behavior of gauge field correlators with finite momentum (k) in a $AdS^4$ black hole background. At low temperature, a substantial non-zero conductivity is only possible for a frequency range $\omega>\omega_g=k$. This tallies with the simple fact that at least an amount of energy $\omega_g$ is needed to create an excitation of momentum $k$. Due to the existence of this ``gap'',one may expect that at zero frequency limit the real part of momentum dependent conductivity falls exponentially with $\frac{1}{T}$. Using analytic methods, we found a $\exp(-\frac{\omega_c}{T})$ falloff of the real part of conductivity with inverse temperature. Interestingly, $\omega_g \neq \omega_c$. From the above results we speculate that the ``degrees of freedoms'', say carriers, different than quasi particle excitation determines conductivity at low temperature and low frequency limit. Here $\omega_c < \omega_g$ and we may calculate their ratios analytically. We also discuss similar issues at a finite chemical potential. Situation is rather different for an extremal blackhole. A zero temperature extremal blackhole does not show a sharp gap for the finite momentum excitations and the real part of conductivity is always non-zero for any non-zero frequency $\omega$. However the real part of conductivity goes to zero at $\omega\to0$ limit. Not surprisingly, we find a powerlaw decay with temperature for the same quantity, as the extremal limit is approached. 

\end{abstract}

\section{Introduction: Carriers vs Quasiparticles}

\par
Gauge-gravity duality\cite{Maldacena:1997re} provides a laboratory to study various field theoretic phenomenon using black holes. Especially the transport properties, including conductivity and viscosity, at finite temperature or chemical potential of strongly coupled model gauge theories may be computed using holographic methods \cite{Policastro:2001yc,Son:2007vk}. These studies involve the calculation of Green's functions in holographic black hole back grounds. Recently, similar study of Green's function for various different holographic models have been used to discuss interesting phenomenon like non-fermi liquid \cite{Liu:2009dm}, superconductivity \cite{Hartnoll:2008vx}.
\par
Here in a simple setup, we look at the physical problem of how a low temperature system responses under finite momentum perturbations . We hope to address the question about the nature of degrees of freedoms responsible for transport at this limit. To illustrate the exact problem we are looking at, let us consider a simple model system, e.g. relativistic particles in a box, and study the response of the system under a space varying disturbance. At zero temperature, to excite a mode with momentum $k$ we have to supply the necessary energy $\omega_g=c_1 k$, determined by the dispersion relation of the system. Hence absorption\footnote{Absorption here broadly refers to source of any dissipative mechanism, where at the first step the system absorbs energy from the incident disturbance. The resulting excited system may then thermalize. } is zero for a frequency less than $\omega_g$ and a non-zero absorption is only possible for $\omega>\omega_g$. The situation is more interesting at a finite temperature($T$). Here the thermal fluctuations may add up with the energy of the incident disturbance to  supply the necessary energy $\omega_g$ needed for absorption. Absorption at $\omega \rightarrow 0$ limit, is purely due to thermal fluctuations. It is natural to expect that for a temperature $T \ll \omega_g$ the absorption ($\sigma$) at zero frequency limit behaves like,
\begin{eqnarray}
 \sigma\sim\exp(-\frac{\omega_g}{T}), \quad \omega \rightarrow 0.
\end{eqnarray}

\begin{figure}
\begin{center}
\includegraphics[scale=0.7]{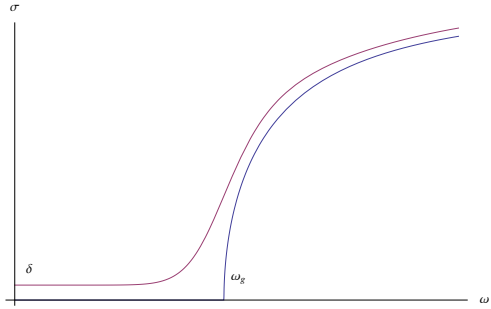}
\end{center}
\caption{Schematic plot of absorption at $T=0$ (lower curve) and at a small non-zero temperature (upper curve). The absorption is significant only for $\omega > \omega_g$, the ``gap'' in the system. However the value of absorption, i.e. $\delta$, at low temperature, low frequency limit scales as $\delta \sim \exp(-\frac{\omega_c}{T})$. Here $\omega_c \neq \omega_g$.} 
\label{fig1}
\end{figure}

However this simplistic description is not true in general. For a complex system the spectra shows quasi particle excitations. In general absorption is only significant if we provide enough energy to excite those quasi particles. However the quasi particles are sometimes made up of more ``fundamental'' degree's of freedoms (say ``carriers''). These carrier degrees of freedoms may not be directly excitable, however they may play a role in conductivity etc. at low temperature.  One familiar example would be a superconductor where the role of "quasi particles" is taken by Cooper pairs. However how the conductivity behaves at low temperature and low frequency limit is determined by the energy of the "carriers" , i.e. electrons. Thermal fluctuations may break a tiny amount of Cooper pairs into electrons and they may take part in the conductivity. Although carriers (electrons) may itself be thought as quasi particles\footnote{The terminology ``carriers'' and ``quasi particles'' used in this paper is just to differentiate between various degrees of freedoms, depending on which are excitable through external perturbation.}, coupling to the electron and photon is such that a incident electromagnetic disturbances may only excite a cooper pair\footnote{For similar phenomenon in the context of different models of holographic superconductivity see \cite{Hartnoll:2008kx,Hartnoll:2008vx,Horowitz:2008bn,Ammon:2008fc,Basu:2008bh}.}.. Hence mass gap perceived by electromagnetic excitation is the energy of the Cooper pairs. Unlike our main consideration in the current work, this particular example does not involve a space varying fluctuation. However, this  demonstrate our main point that the energy gap calculated at zero temperature from absorption may not same as the energy gap calculated from the zero frequency zero temperature limit of the absorption (see Fig \ref{fig1}). 

In our case, i.e. with a non-zero momentum $k$, we may expect that quasi particles excitations and carriers will show different dispersion relation. Lets assume that for a temperature $T \ll \omega_g$ the absorption ($\sigma$) at zero frequency limit behaves like,
\begin{eqnarray}
 \lim_{\omega \rightarrow 0} \sigma\sim\exp(-\frac{\omega_c}{T}), 
\label{wc1}
\end{eqnarray}
On general ground we expect $\omega_c \neq \omega_g$. In most cases we expect $\frac{\omega_c}{\omega_g} < 1$. To demonstrate these ideas in a holographic context, we will choose a simple system of a Schwarzschild black hole in $AdS_4$ and study the transverse component of the gauge field correlators in this back ground. By studying conductivity, we can analytically calculate $\omega_g$ and $\omega_c$ taking a zero temperature/large momentum limit. We argue that,
\begin{eqnarray}
\frac{\omega_c}{\omega_g}= \frac{3}{2\sqrt{\pi }}\frac{ \Gamma \left(\frac{4}{3}\right)}{\Gamma \left(\frac{5}{6}\right)}\approx 0.669. 
\end{eqnarray}

  In the language of dual field theory we are studying U(1) (e.g. R-charge, baryon number etc) current correlators. Depending on the particulars of the dual theories, these current operators would be various bilinears in elementary fields. In short they could be thought as a mesonic operator consisted of more elementary ``quarks''. These ``quarks'' possibly play the role of carrier in our picture. Although we may not directly calculate the co-correlators of the ``quark'' fields from the holographic geometry, their effects may be visible in mesonic co-correlators. It should also be kept into mind that what we call carrier and quasi particles may itself depend on the operator whose correlation we are looking at. In some sense carries are also a type of quasi particles. The exact situation is not clear to us at this moment. 

  We also discuss similar gap calculation in the presence of a non-zero chemical potential. In the canonical ensemble the ratio $\frac{\omega_c}{\omega_g}$ decreases with charge $q$ and goes to zero in the extremal limit. In grand canonical ensemble the system approaches an extremal black hole as the temperature is lowered. Although of zero temperature, the extremal black geometry does not show a sharp gap for finite momentum excitations and the real part of conductivity is non-zero for any non-zero frequency $\omega$. However the real part of conductivity goes to zero at $\omega\to0$ for a non-zero $k$. Which may be considerd as a type of ``weak gap''. As the temperature is lowered from an non-extremal solution the real part of the conductivity at zero temperature limit decays as a power law in $T$,
\begin{eqnarray}
 \lim_{\omega \rightarrow 0} \sigma\sim T^{\gamma}, \quad \gamma>0
\label{wc2}
\end{eqnarray}
 This is in cotrast to \eref{wc1}, which shows an exponential decay.
\section{Setup}
Let us consider a Schwarzschild black hole in $AdS^4$,
\begin{equation}
 ds^2 = L^2 \alpha^2(-f(z) dt^2  +  \frac{1} {z^2}  (dx^2 + dy^2) ) + \frac{L^2 dz^2}{z^4 f(z)}.
\label{bkgr}
\end{equation}
with
\begin{equation}
f(z) =\frac{1}{z ^2}-z.
\label{metric}
\end{equation}
The dual boundary theory of such a setup would be $2+1$ dim CFT at finite temperature.
The black hole horizon is located at $z_h=1$ and boundary of AdS is placed at $z=0$. The parameter $L$ is the radius of the $AdS^4$ space. $\alpha$ is related to the black hole mass and has the dimension of inverse length. The temperature of the black hole is given by,
\begin{equation}
 T =\frac{3\alpha}{4\pi }
\label{temp}
\end{equation}
 by rescaling 
\begin{eqnarray}
t \rightarrow t/\alpha,\quad x \rightarrow x/\alpha,\quad y \rightarrow y/\alpha
\label{rescale}
\end{eqnarray}
one can set $\alpha=1$.

In the background (\ref{bkgr}) we will consider a Abelian gauge field given by the action, 
\begin{eqnarray}
S =-\int dx^4 \sqrt{-g} \frac{L^2}{4}F^{a b}F_{ab}
\end{eqnarray}
We will not consider the gravity back reaction of the gauge fields.  Due to the transnational invariance of the boundary coordinates $t,x,y$, the gauge potential can be Fourier transformed,
\begin{eqnarray}
\hat{A} (z,t,x,y)= \int d^4 x  \hat A(z,\hat{k}) \exp(i\omega t + i\vec{k}\cdot \vec{x}). 
\end{eqnarray}

Without loss of generality one may we choose the $\hat {k}=(\omega,0,k)$. We will only consider the transverse component $A^{(\omega,k)}_x$ which decouples from the other components \cite{Herzog:2007ij}. The equation of motion for $A^{(\omega,k)}_x$ is,  
\begin{eqnarray}
 A^{''(\omega,k)}_x &+&  \left ( \frac{ 2}{ z} + \frac{  f'}{  f }  \right) A^{'(\omega,k)}_x+\frac{\omega^2}{z^4 f^2 } A^{(\omega,k)}_x -\frac{  k^2 }{ z^2 f } A^{(\omega,k)}_x =0.
\label{maineq}
\end{eqnarray}
The asymptotic behavior of $A^{(\omega,k)}_x$ as $z\rightarrow 0$ is given by, 
\begin{eqnarray}
A^{(\omega,k)}_x (z) \sim S^{(\omega,k)}_x +z J^{(\omega,k)}_x+ \cdots
\end{eqnarray}
where $S^{(\omega,k)}_x$ is the boundary value of gauge potential and $J^{(\omega,k)}_x$  is the boundary current\footnote{All quantities are defined in momentum space. For simplicity we will discard the $\omega,k$ indices from now on.}. Under the rescaling (\ref{rescale}), we have 
\begin{eqnarray}
 \omega \rightarrow \alpha \omega,\quad k \rightarrow \alpha k,\quad S_x \rightarrow \alpha S_x,\quad J_x \rightarrow \alpha^2 J_x. 
\end{eqnarray}
Hence $\frac{\omega}{T},\frac{k}{T},\frac{S_x}{T}$ and $\frac{J_x}{T^2}$ may be thought as the dimensionless quantities.
To take the small temperature limit we can fix $T$ and take a large $k$ limit\footnote{This is just opposite to the more familiar $k \rightarrow 0$ fluid dynamics limit, e.g. in \cite{Policastro:2002se,Bhattacharyya:2008jc}.}. We will take this approach in the current paper. 

\section{Conductivity and gap calculation}

Frequency dependence of the gauge field response functions have been discussed in \cite{Herzog:2007ij}. Many interesting properties are determined by the electro magnetic dualities. Here we will try to solve the equations explicitly. The \eref{maineq} may be written as,
\begin{eqnarray}\label{seceq}
z^2 f (z^2 f A_x')'+ \omega^2 - k^2 z^2 f A_x&=&0 \\
\nn \frac{d^2}{d\tilde z^{2}} A_x + \omega^2 A_x -k^2 h(\tilde z)  A_x&=&0,\quad  d\tilde z=\frac{dz}{z^2 f}, \quad h(\tilde z)=z^2 f(z).
\end{eqnarray}
and near the boundary $\tilde z \rightarrow 0$ and near the black hole horizon $\tilde z \rightarrow \infty$. The potential $h(\tilde z)$ vanishes at the horizon and is unity at the boundary. In the near horizon region, the equation \eref{seceq} has two linearly independent solution given by, $A_x \sim e^{\pm i \omega \tilde z}$. Following the standard prescription of the finite temperature holography\cite{Son:2007vk}, we will use the ingoing wave boundary condition (i.e. $A_x \sim A^{H}_x e^{-i\omega \tilde z}$) at the horizon for our calculations. 

The conductivity is given by,
\begin{eqnarray}
\sigma(k,\omega)=\frac{G^{R}(k,\omega)}{i \omega}=\frac{J_x}{i\omega S_x} 
\label{cond}
\end{eqnarray}
where $G_R$ is the retarded Green's function. At zero momentum, the conductivity is a real constant independent of frequency ,
\begin{equation}
 \sigma(0,\omega)= 1 
\end{equation}

However at finite $k$, the conductivity varies with frequency $\omega$ and no-longer a real number. We will concentrate on the real part of the conductivity which by \eref{cond} is related to the imaginary part of Green's function and consequently dissipation. At large $k$, we may scale away the black hole horizon by defining $z\rightarrow \frac{z}{k}$ and get back the results of $AdS^4$ space. Hence at large $k$ the leading part of the conductivity is,
\begin{eqnarray}
\text{Re}(\sigma)&\approx& \frac{1}{\omega}\sqrt{\omega^2-k^2},\quad \omega > k \\
\nn &\approx& 0, \quad \omega < k
\label{res1}
\end{eqnarray}

This suggest a value $\omega_g=k$. This is consistent with Lorentz symmetry of the emergent $AdS^4$ space. This is sufficient for our purpose. However by a careful numerical and semi-analytic calculations one may show that for large omega there is a quasi normal mode at $\omega \approx k - i c_1$, where $c_1$ is a constant. Actually at large $k$, all the quasi normal poles line up near the line $\omega \approx \pm k - \text{imaginary}$ in the lower half of complex $\omega$ plane. Although the gap between poles do not vanish in the complex $\omega$ plane, but they may be thought to produce a branch cut in $\frac{\omega}{k}$ plane \cite{pallab:un}. This is consistent with the appearance of a square root in \eref{res1}.
   
Next, we proceed to calculate $\omega_c$ and check whether it is same as $\omega_g$. As discussed $\omega_c$ is calculated from the large momentum behavior of the real part of the conductivity at zero momentum limit. From \eref{seceq} we get,
\begin{eqnarray}
\frac{d}{d\tilde z} \text{Im}(A_x^{*} \frac{d}{d\tilde z} A_x)=0.
\end{eqnarray}

Integrating the above equation from horizon to the boundary, we get
\begin{eqnarray}
\text{Im}(S_x^{*} J_x) &=& \text{Im}(A^{H*}_x\frac{d}{d\tilde z} A^{H})=\text{Im}(i \omega A^{H*}_x A^{H}_x) \\
\Rightarrow  \text{Re}(\sigma)&=&\text{Re}(\frac{J_x}{i \omega S_x})= \frac{|A^{H}_x|^2}{|S_x|^2}
\label{condfor}
\end{eqnarray}
Here $A^{H}_x$ is the amplitude of the gauge field near the horizon. We like to understand the zero frequency limit of \eref{condfor} more carefully. Strictly at $\omega=0$, \eref{condfor} is not valid and $\text{Re}(\sigma$) has a $\delta$ function divergence. This divergence comes from the existence of dissipationless magnetization current. However $\lim_{\omega\to0} \text{Re}(\sigma)$ is well defined and finite. This limiting quantity provides the information about the dissipation in our system. At $\omega \to 0$ limit, the ingoing boundary condition at the horizon is replaced by the regularity condition at the horizon ($z=1$),
\begin{eqnarray}
 A_x'(1)=\frac{k^2}{f'(1)} A_x(1).
\end{eqnarray}
 Using the above boundary condition we solve \eref{maineq} with $\omega=0$ to get the ratio in \eref{condfor}. Furthermore at  $k \rightarrow \infty$ limit the WKB method gives an exact solutions to the equations \eref{seceq}. The real part of the conductivity is then calculated as, \footnote{A similar formula for $\Delta$ has been calculated in the context of scalar propagators in \cite{Hartnoll:2008hs}. I thank Sean Hartnoll for pointing this out. However the apparent mismatch between WKB and numerical method is possibly a small $k$ artifact. },
\begin{eqnarray}
 \text{Re}(\sigma) &\sim& \exp(-2 k \Delta), \\
\Delta &=& \int_0^1 \frac{dz}{z\sqrt{f(z)}} = \frac{\sqrt{\pi } \Gamma \left(\frac{4}{3}\right)}{\Gamma \left(\frac{5}{6}\right)}.
\label{decay}
\end{eqnarray}

Using \eref{temp}, we may calculate $\omega_c=2kT\Delta=2 k \frac{3}{4 \pi} \Delta \approx 0.669 k$. As we expected, $\frac{\omega_c}{k} < 1$. It is interesting note the near rational ratio  $\frac{\omega_c}{\omega_g}\approx 0.669 \approx \frac{2}{3}$.

\section{The case with non-zero chemical potential}

With a non-zero chemical potential the metric function in \eref{bkgr} changes to \cite{Romans:1991nq},
\begin{eqnarray}
f_q(z)=\frac{q^2 z^4-\left(q^2+1\right) z^3+1}{z^2}
\end{eqnarray}
The background gauge field is given by \footnote{This gauge field $A_t$ does not have to be the component of same gauge field whose fluctuations we are looking at.},
\begin{eqnarray}
A_t=2q\alpha(z-1).
\end{eqnarray}
The boundary value of the gauge field, i.e. chemical potential, is given by $\mu=2 q \alpha$. The temperature of the black hole is given by,
\begin{eqnarray}
T=\frac{\alpha}{4\pi}(3-q^2)
\end{eqnarray}
\subsection*{Canonical ensemble}
\begin{figure}
\begin{center}
\includegraphics[scale=0.5]{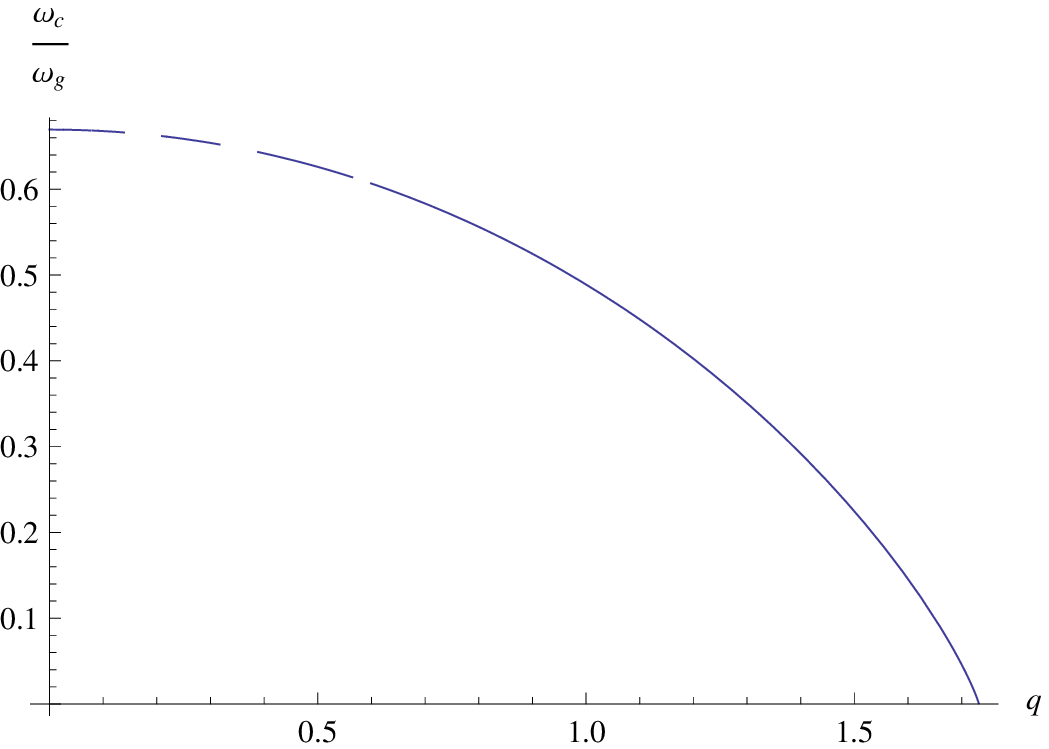}
\hspace{0.3cm}
\includegraphics[scale=0.5]{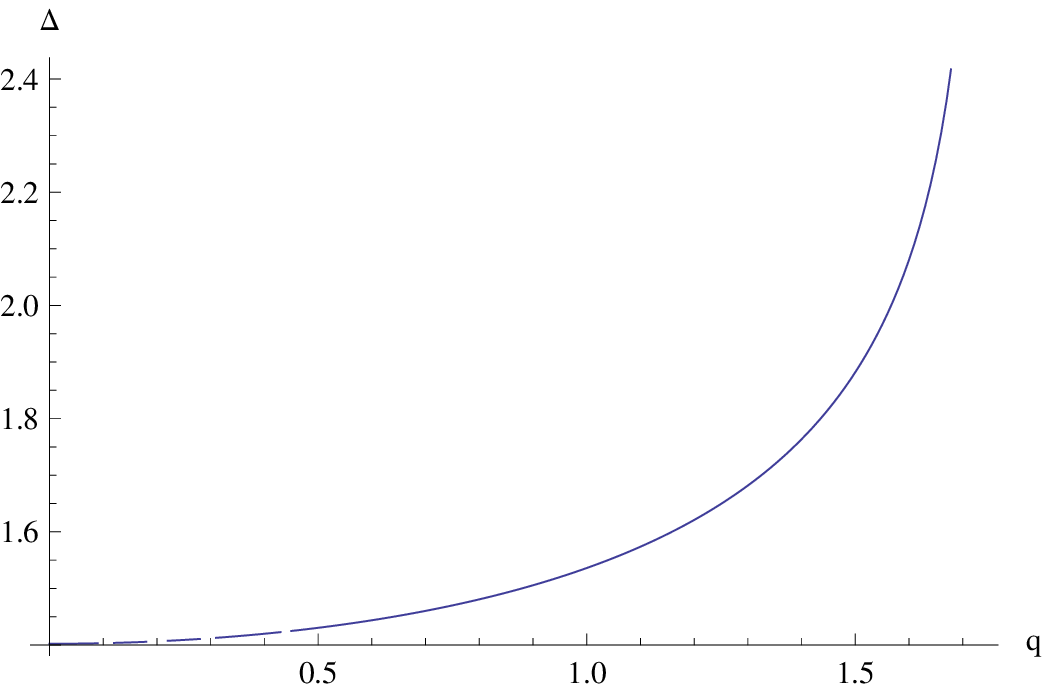}
\end{center}
\caption{The curve on the left hand side shows how $\frac{\omega_c}{\omega_g}$ varies as $q$ is varied. The righthand one is for $\Delta$ vs $q$.} 
\label{fig2}
\end{figure}

Here taking a large momentum limit like in the previous section, results in a system with $T\rightarrow 0$ and $\frac{\mu}{T}$(or equivalently $q$) kept fixed. Here also correlators decay at large $k$ with a $\Delta_q$ (similar to $\Delta$ in \eref{decay},
\begin{eqnarray}
\Delta_q=\int_0^1 \frac{dz}{z\sqrt{f_q(z)}}.
\end{eqnarray}
The above integral may be evaluated in terms of elliptic integrals, however we wont write down the explicit formula here. One can numerically plot the value of $\Delta_q$ (see Fig\ref{fig2}). In the extremal limit $q \rightarrow \sqrt{3}$ and $\Delta_q$ diverges as,
\begin{eqnarray}
 \Delta_q \sim -\log(3-q^2).
\end{eqnarray}
From $\Delta_q$ one may calculate $\frac{\omega_c}{k} = 2 T \Delta=\frac{3-q^2}{4 \pi} \Delta$. 
 However in the extremal limit,${\omega_c}$  tends to zero as,
\begin{eqnarray}
 \frac{\omega_c}{k} \sim -(3-q^2)\log(3-q^2).
\end{eqnarray}
Chemical potential does not affect the value of $\omega_g=k$. The ratio $\frac{\omega_g}{\omega_c}(=\frac{\omega_g}{k})$ is plotted in Fig \ref{fig2}. 

\begin{figure}
 \begin{center}
 \includegraphics[scale=0.7]{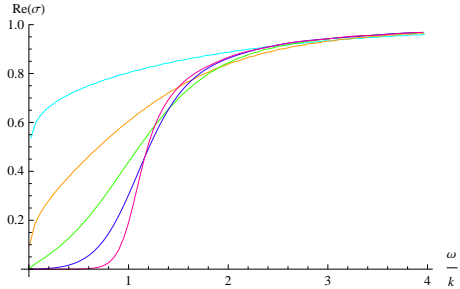}
\end{center}
\caption{Plot of $\mathrm{Re}(k,\sigma)$ with $\frac{\omega}{k}$. From left $k=0.5,1,\sqrt(3),2,6$.}
\label{fig:plotsigma}
\end{figure}

\subsection*{Grandcanonical ensemble}
One may also investigate what happens to the conductivity as $T \rightarrow 0$ with $\mu$ (instead of $\frac{\mu}{T}$) kept fixed.
In this limit the black hole geometry tends to an extremal one. For an extremal black hole, the near horizon geometry is $AdS^2 \times R^2$. In this near horizon geometry one may solve the \eref{maineq} exactly in terms of Hankel functions \cite{Horowitz:2009ij}. Following the methods of matched asymptotic expansion it is not difficult to show,
\begin{equation}
 \text{Re} (\sigma(k,\omega)) \sim \omega^{\sqrt{1+k^2}-1}
\end{equation}
 for small $\omega$. However for a non-zero value of $\omega$, $\text{Re} (\sigma(k,\omega))$ is always non-zero (see Fig \ref{fig:plotsigma}) as the potential $h(\tilde z)$ in \eref{seceq} vanishes near to horizon and of finite height. Hence the ratio between the strength of the gauge field at horizon and boundary remains finite and the real part of the conductivity is non-zero. It is interesting that evenif the system is of zero temperature, there is no hard gap for modes with a finite momentum. This fact is possibly related to large-N limit and huge degeneracy of states for an extremal ground state. Due to the emergent lorentz invariance, at a very large value of $k$, $\text{Re} (\sigma)$ is negligible in the region $\omega < k$. 

One may investigate how the real part of conductivity decays at zero frequency limit with temperature $T$. That is we want to find out the low temperature behavior of the quantity $\lim_{\omega\to0} \text{Re}(\sigma(k,\omega))$. Due to the absence of a ``hard gap'' in the extremal limit, it is natural to expect a power-law falloff with $T$. Let us look at the near horizon region such that $r=1-z \ll 1$. The metric function may be expanded as a power series in the near horizon co-ordinate $r$ as,
\begin{equation}
 f(r)\approx c r+d r^2 + o(r^3).
\end{equation}
where $c=(3-q^2) \propto T$ and $d=3+q^2$. In the extremal limit $c\to0$ and $d\to6$. In the near horizon limit the regular(i.e. regular at the horizon) solution of the equation \eref{maineq} is given by,
\begin{equation}
 A(r) \sim (c+d r)^{\frac{k^2}{d}}.
\end{equation}
This exact solution is valid in a region $ r \ll 1$, where as a perturbation in $c$ is valid in a region $r \gg \frac{c}{d}$. As $\frac{c}{d} \ll 1$, we can have a intermediate region $\frac{c}{d} \ll r_* \ll 1$ where both of the condition is valid. Using the method of matched asymptotic expansion we can calculate the real part of conductivity as,
\begin{eqnarray}
 \lim_{\omega\to0}\text{Re}(\sigma) = \left(\frac{A_H}{A_x(r_*)}\frac{A_x(r_*)}{S_x}\right)^2 \sim c^\frac{2k^2}{d} \sim T^\frac{k^2}{3}=T^\frac{4 k^2}{\mu^2}.
\end{eqnarray}

\section{Discussion and Future directions}

It would be interesting to understand the physical mechanism behind our results in more detail. It should be noted that $\omega_g$ is determined by the behavior of the potential $h(\tilde z)$ near the boundary $\tilde z=0$. Whereas $\omega_c$ is sensitive to the whole range of $\tilde z$. Near the boundary $h(\tilde z)$ is $1$ and the Lorentz symmetry is restored. This fixes $\omega_g$ to $k$. However there is no such constraint on $\omega_c$ and generically one would expect $\omega_c<k$ from causality. At a technical level, it would be nice to have more through understanding of the poles and zeros of retarded green's function in complex $\omega$ plane. 

To get a clearer picture it would be interesting to repeat similar gap calculations with different fields, including scalars and fermionic fields, and different geometries including non-conformal ones \cite{Horowitz:1998ha} and brane systems \cite{Karch:2002sh}.

\section*{Acknowledgments}
I like to thank Mark Van Raamsdonk, Hsien-Hang Shieh, Anindya Mukherjee, Jianyang He for discussion. I like to thank all members of UBC string theory department for their support and encouragement. I also thank Hong Liu, Gary Horowitz and especially Sean Hartnoll for comments. I acknowledge support from the Natural Sciences and Engineering Research Council of Canada.

\bibliographystyle{hunsrt}
\bibliography{car}

\end{document}